\documentclass{appolb}

\usepackage{amssymb}
\usepackage{epsfig}
\usepackage{eucal}
\usepackage{amsmath}

\newcommand{\non}{\nonumber}
         
\newcommand{\Eq}[1]{Eq.(\ref{#1})}
\newcommand{\be}{\beta}

\begin{document}
\pagestyle{plain}
\title{RADIATIVE RETURN AT NLO AND THE \\ MEASUREMENT OF THE  
HADRONIC CROSS-SECTION
\thanks{Presented at the XXV International Conference 
on Theoretical Physics ``Particle Physics and Astrophysics in the 
Standard Model and Beyond'', Ustro\'n, Poland, 9-16 September 2001.}}
\author{Germ\'an Rodrigo
\address{Institut f\"ur Theoretische Teilchenphysik,
Universit\"at Karlsruhe, \\ D-76128 Karlsruhe, Germany. \\
TH-Division, CERN, CH-1211 Gen\`eve 23, Switzerland.\\
e-mail: \tt{German.Rodrigo@cern.ch}}
}
\maketitle
\vspace{-6.5cm}
\hfill TTP01-26
\vspace{5.5cm}

\begin{abstract}
The measurement of the hadronic cross-section in $e^+ e^-$
annihilation at high luminosity factories using the radiative 
return method is motivated and discussed. A Monte Carlo 
generator which simulates the radiative process 
$e^+ e^- \rightarrow \gamma+hadrons$ at the next-to-leading
order accuracy is presented. The analysis is then extended to the
description of events with hard photons radiated at very small angle. 
\end{abstract}
\PACS{13.40.Em, 13.40.Ks, 13.65.+i}

\section{Motivation}

Electroweak precision measurements in present particle physics provide 
a basic issue for the consistency tests of the Standard Model (SM) or 
its extensions. New phenomena physics can affect low energy processes 
through quantum fluctuations (loop corrections). Deviations from the 
SM predictions can therefore supply indirect information about new 
undiscovered particles or interactions. 

The recent measurement of the muon anomalous magnetic moment 
$a_{\mu} \equiv (g-2)_{\mu}/2$ at BNL~\cite{Brown:2001mg} reported a 
new world average showing a discrepancy at the $2.6$ standard deviation 
level with respect to the theoretical SM evaluation of the same quantity
which has been taken as an indication of new physics.
For the correct interpretation of experimental data the appropriate 
inclusion of higher order effects as well as a very precise knowledge 
of the SM input parameters is required. The BNL experiment plans a new 
measurement with an accuracy three times smaller which will challenge 
even more the theoretical predictions. 

One of the main ingredients of the theoretical prediction for the 
muon anomalous magnetic moment is the hadronic vacuum polarization 
contribution which moreover is responsible for the bulk of the 
theoretical error. This quantity plays also an important role in the 
evolution of the electromagnetic coupling $\alpha_{QED}$ from the 
low energy Thompson limit to high energies. In both cases
the precise knowledge of the ratio
\begin{equation}
R(s)=\frac{\sigma(e^+e^+ \rightarrow hadrons)}
{\sigma(e^+e^-\rightarrow\mu^+ \mu^-)}~,
\end{equation}
over a wide range of energies is required. The measurement of 
the hadronic cross section in $e^+ e^-$ annihilation to an 
accuracy better than $1\%$ in the energy range below $2$~GeV 
is necessary to improve the accuracy of the present predictions 
for the muon anomalous magnetic moment and the QED coupling. 

In this paper, the radiative return method is motivated and 
described. It has the advantage against the conventional 
energy scan~\cite{Akhmetshin:1999uj}, that the systematics of the 
measurement (e.g. normalization, beam energy) have to be taken into 
account only once but not for each individual energy point independently. 
An improved Monte Carlo generator which simulates the radiative 
process $e^+e^- \rightarrow \gamma + hadrons$ at the Next-to-Leading
Order (NLO) accuracy is presented. Finally, the analysis is extended 
to the description of events with hard photons radiated at very small 
angle.

\section{The muon anomalous magnetic moment and the QED coupling}

\begin{figure}
\begin{center}
\epsfig{file=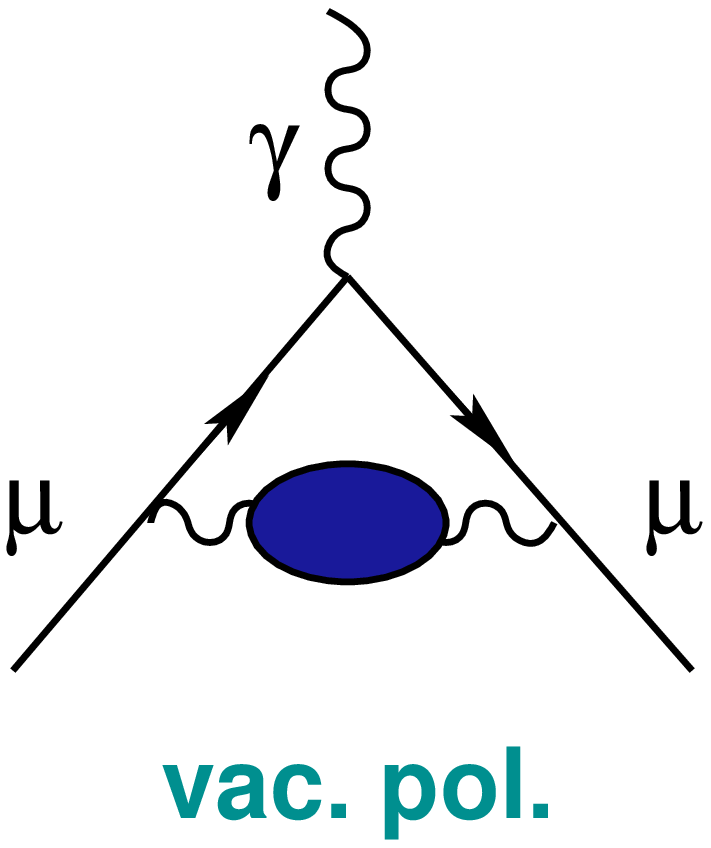,width=3.cm} $\qquad$
\epsfig{file=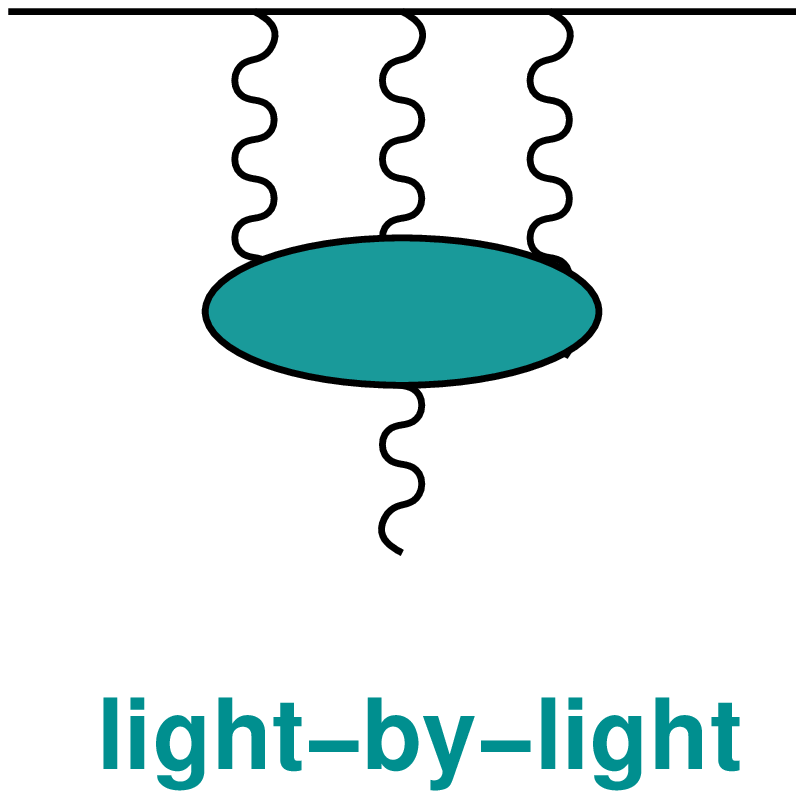,width=3.4cm}
\end{center}
\caption{Vacuum polarization and light-by-light hadronic contributions
to the muon anomalous magnetic moment.}
\label{fig:muon}
\end{figure}

The Standard Model prediction for the muon anomalous magnetic moment
consist of three contributions~\footnote{For recent reviews 
see~\cite{reviews}.}
\begin{equation}
a_{\mu} = a_{\mu}^{\mathtt{QED}} + a_{\mu}^{\mathtt{had}}
+ a_{\mu}^{\mathtt{weak}}~.
\end{equation}
The QED correction, known (estimated) up to five loops~\cite{amuQED}, 
gives the main contribution (see Table~\ref{tab:amu} for numerical 
values~\footnote{N.A. see~\cite{newlight} 
for a recent evaluation of the hadronic light-by-light contribution.}). 
The weak contribution is currently known up to two loops~\cite{amuweak}. 
Finally, the hadronic contribution consist of three different 
terms where the leading one comes from the vacuum polarization 
diagram in figure~\ref{fig:muon} which furthermore is responsible for 
the bulk of the theoretical uncertainty.

\begin{table}
\begin{center}
\begin{tabular}{lrrl}
\hline \hline
& $a_{\mu} \times 10^{11}$ & & \\ \hline
QED  & 116584706 $\pm$ &   3 & \\ 
hadronic ($\mathsf{vacuum \;polarization}$)  &    6924 $\pm$ &  62 &
 \cite{Davier:1998si} \\
hadronic ($\mathsf{light-by-light}$) &  -85 $\pm$ &  25 & \cite{amulight} \\
hadronic ($\mathsf{other \; higher \; orders}$) &   -101 $\pm$ &   6 & 
\cite{Krause:1997rf} \\
weak &                152 $\pm$ &   4 &  \\ \hline \hline
\end{tabular}
\caption{QED, hadronic and weak contributions to the muon anomalous 
magnetic moment.}
\label{tab:amu}
\end{center}
\end{table}

While the QED and the weak contributions are well described 
perturbatively, the hadronic contribution cannot be completely 
calculated from perturbative QCD. The hadronic vacuum polarization 
contribution~\cite{Eidelman:1995ny,Brown:1996eq,Davier:1998si,Narison:2001jt,Jegerlehner:2001wq,DeTroconiz:2001wt}
can however be evaluated from the following dispersion relation  
\begin{equation} 
\quad a_{\mu}^{\mathtt{had}}(\mathsf{vac.pol.}) 
= \left( \frac{\alpha m_{\mu}}{3 \pi} \right)^2 \int_{4m_{\pi}^2}^{\infty}  
\frac{ds}{s^2} \; K(s) \; R(s)~,
\end{equation} 
where the kernel $K(s)$ is a known smooth bounded function
and $R(s)$ is the hadronic ratio which has to be extracted from 
experimental data. Under the assumption of conserved vector current
and isospin symmetry $\tau$ decays can be included also for the 
evaluation of the dispersion integral. 
Due to the $1/s^2$ dependence of the dispersion integral,
the low energy region contributes dominantly to this integral: 
$70\%$ of the result comes from the $\pi \pi$ channel and,
even more, $90\%$ of it is given by the region below $1.8$~GeV.
The precise measurement of the hadronic cross section in $e^+ e^-$
annihilation, specially at low energies, has therefore a crucial 
importance for the accurate determination of the muon anomalous 
magnetic moment. 

The running of the QED fine structure constant
from the Thompson limit to high energies is given by 
\begin{equation}
\alpha(s) = \frac{\displaystyle \alpha(0)}{\displaystyle 1-\Delta \alpha(s)}~,
\qquad
\Delta \alpha = \Delta \alpha_{\mathtt{lep}} + \Delta \alpha_{\mathtt{top}}
+ \Delta \alpha_{\mathtt{had}}~.
\end{equation}
Again, the Standard Model prediction consists of several 
contributions. Among them, the hadronic one which can be 
evaluated from the dispersion relation
\begin{align}
\Delta \alpha_{\mathtt{had}}(s) = -\frac{\alpha s}{3 \pi} Re
\int_{4m_{\pi}^2}^{\infty} & 
\frac{ds'}{s'} \; \frac{R(s')}{s-s'-i \eta}~.
\end{align}
The leptonic~\cite{Steinhauser:1998rq} and the top quark~\cite{top} 
contributions are well described perturbatively while the hadronic 
contribution~\cite{Burkhardt:2001xp} gives the main 
theoretical error (see Table~\ref{tab:alpha} for numerical values). 
The dispersion integral grows in this case only as $1/s$ which 
means that also the high energy points become relevant.
The region below $1.8$~GeV contributes only to $20\%$ of the integral. 
A better knowledge of the hadronic cross section in this region would 
be nevertheless also important to reduce the error. 

\begin{table}
\begin{center}
\begin{tabular}{lrr}
\hline \hline
& $\Delta \alpha (\mathsf{M_Z}) \times 10^{4}$ & \\ \hline
leptonic $\qquad \qquad$ &  314.98 $\; \; \;$  \\ 
top quark &   -0.70 $\pm$ & 0.05   \\ 
hadronic  &  276.1  $\,$ $\pm$ &  3.6 $\,$  \\ \hline \hline
\end{tabular}
\caption{Leptonic, top quark and hadronic contributions to the  
running of the QED fine structure constant at $M_Z$.}
\label{tab:alpha}
\end{center}
\end{table}

\section{The radiative return method and the hadronic cross section}

\begin{figure}
\begin{center}
\epsfig{file=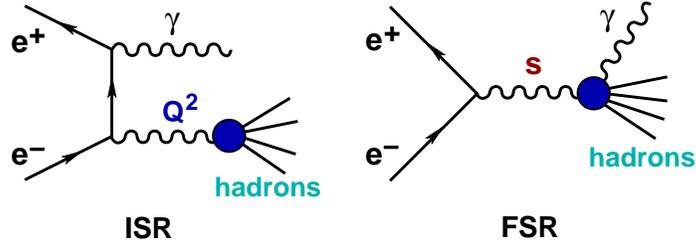,width=10cm}
\end{center}
\caption{Initial state radiation (ISR) and final state radiation (FSR) 
in the annihilation process $e^+ e^- \rightarrow \gamma + hadrons$.}
\label{fig:isrfsr}
\end{figure}

\begin{figure}
\begin{center}
\epsfig{file=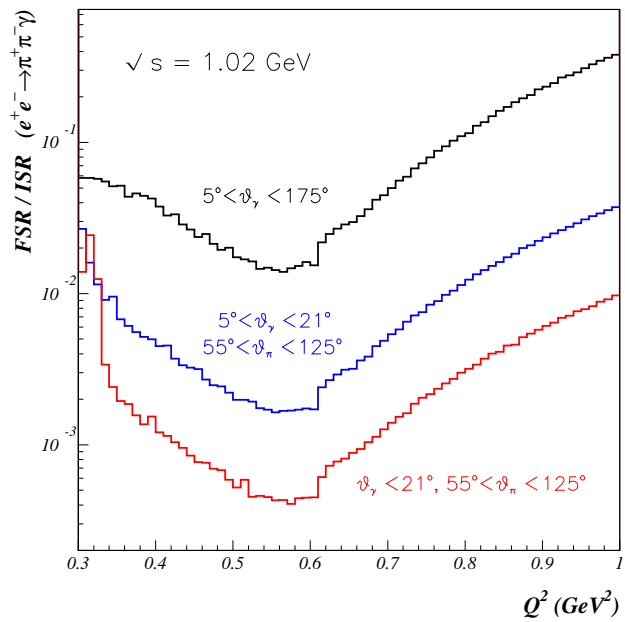,width=9cm}
\end{center}
\caption{Relative contribution of FSR versus ISR in the annihilation
process $e^+ e^- \rightarrow \pi^+ \pi^- \gamma$ at $\sqrt{s}=1.02$GeV
for different photon and pion angular cuts. $E_{\gamma}>10$MeV.}
\label{fig:eva}
\end{figure}

The radiative process $e^+ e^- \rightarrow \gamma + hadrons$, where the 
photon is radiated from the initial particles (initial state radiation, ISR),
see figure~\ref{fig:isrfsr}, can be used to measure the hadronic cross section 
$\sigma(e^+e^-\rightarrow hadrons)$ at high luminosity electron-positron 
storage rings, like the $\phi$-factory DAPHNE or at $B$-factories, over 
a wide range of energies. The radiated photon reduces the effective energy
of the collision and thus the invariant mass of the hadronic system. 
This possibility has been proposed and studied in detail 
in~\cite{Binner:1999bt} (See also~\cite{Spagnolo:1999mt}).
A Monte Carlo generator called EVA~\cite{Binner:1999bt} which simulates 
the process $e^+ e^- \rightarrow \pi^+ \pi^- \gamma$ was built.
The four pion channel was considered in~\cite{Czyz:2000wh}.

Radiation of photons from the hadronic system (final state radiation, FSR) 
should be considered as the background of the measurement. 
One of the main issues of the radiative return method is the 
suppression of this kind of events by choosing suitable kinematical cuts. 
From EVA studies this is achieved by selecting events with the tagged photons 
close to the beam axis and well separated from the hadrons which reduces 
FSR to a reasonable limit. This is illustrated in figure~\ref{fig:eva}.
Furthermore, the suppression of FSR overcomes the problem of its model 
dependence which should be taken into account in a completely inclusive 
measurement~\cite{Hoefer:2001mx}.

Preliminary experimental results using the radiative return method have 
been presented recently by the KLOE collaboration~\cite{Adinolfi:2000fv}.
Large event rates were also observed at BaBar~\cite{babar}.

The theoretical description of the radiative events under 
consideration to a precision better than $1\%$ requires a 
precise control of higher order radiative corrections. 
The next sections are devoted to the calculation of the 
NLO corrections to ISR and its inclusion in a new improved 
Monte Carlo generator~\cite{Rodrigo:2001jr,inpreparation}.

\section{NLO corrections to ISR}

\begin{figure}
\begin{center}
\epsfig{file=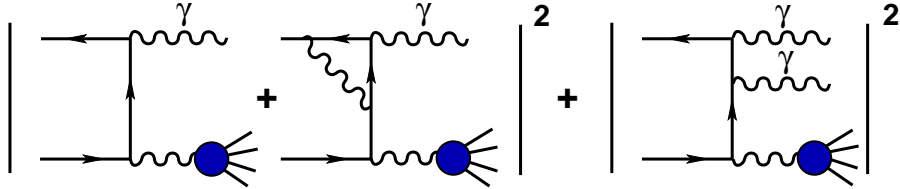,width=12cm}
\end{center}
\caption{NLO corrections to initial state radiation in the annihilation 
process $e^+ e^- \rightarrow \gamma + hadrons$.}
\label{fig:nlo}
\end{figure}

At NLO, the $e^+ e^-$ annihilation process
\begin{align}
e^+(p_1) + e^-(p_2) \rightarrow & \gamma^*(Q) + \gamma(k_1)~,
\end{align}
where the virtual photon decays into a hadronic final state,
$\gamma^*(Q) \rightarrow$ hadrons, and the real one is emitted
from the initial electron or positron, receives contributions
from one-loop corrections and from the emission of a second real
photon (see figure~\ref{fig:nlo} for a schematic representation).

The calculation of the different contributions proceeds as follows:
first the one-loop diagrams are reduced, using the standard 
Passarino-Veltman procedure~\cite{Passarino:1979jh}, to the calculation of 
a few one-loop scalar integrals which are calculated in $D=4-2\epsilon$ 
dimensions in order to regularize their divergences. The interference with 
the lowest order Feynman diagrams is calculated and the ultraviolet (UV) 
divergences are renormalized using the on-shell mass scheme. 
The remaining infrared (IR) divergences are cancelled
by adding the soft contribution of the second photon 
calculated analytically by integration in phase space 
up to a soft energy cutoff $E_{\gamma}<w\sqrt{s}$ far below 
the center of mass energy $\sqrt{s}$. 

In order to facilitate the extension of the Monte Carlo simulation 
to different hadronic exclusive channels the differential rate is cast 
into the product of a leptonic and a hadronic tensor and the corresponding 
factorized phase space
\begin{equation}
d\sigma = \frac{1}{2s} L_{\mu \nu} H^{\mu \nu}
d \Phi_2(p_1,p_2;Q,k_1) d \Phi_n(Q;q_1,\cdot,q_n) 
\frac{dQ^2}{2\pi}~,
\end{equation}
where $d \Phi_n(Q;q_1,\cdot,q_n)$ denotes the hadronic 
$n$-body phase-space including all statistical factors 
and $Q^2$ is the invariant mass of the hadronic system.
The physics of the hadronic system, whose description is model 
dependent, enters only through the hadronic tensor $H^{\mu \nu}$.

The leptonic tensor which describes the next-to-leading order 
virtual and soft QED corrections to initial state radiation in 
$e^+ e^-$ annihilation has the following general form: 
\begin{align}
L^{\mu \nu}_{\mathrm{virt+soft}} &=
\frac{(4 \pi \alpha)^2}{Q^4 \; y_1 \; y_2} \;
\bigg[ a_{00} \; g^{\mu \nu} + a_{11} \; \frac{p_1^{\mu} p_1^{\nu}}{s}
 + a_{22} \; \frac{p_2^{\mu} p_2^{\nu}}{s} \non \\
&+ a_{12} \; \frac{p_1^{\mu} p_2^{\nu} + p_2^{\mu} p_1^{\nu}}{s}
+ i \pi \; a_{-1} \; 
\frac{p_1^{\mu} p_2^{\nu} - p_2^{\mu} p_1^{\nu}}{s} \bigg]~,
\label{generaltensor}
\end{align}
where $y_i=2k_1\cdot p_i/s$ with $p_1$ ($p_2$) the four momentum of the 
positron (electron).
The scalar coefficients $a_{ij}$ and $a_{-1}$ allow the following expansion 
\begin{equation}
a_{ij} = a_{ij}^{(0)} + \frac{\alpha}{\pi} \; a_{ij}^{(1)}~, \qquad
a_{-1} = \frac{\alpha}{\pi} \; a_{-1}^{(1)}~,
\end{equation}
where $a_{ij}^{(0)}$ give the LO contribution. 
The NLO coefficients $a_{ij}^{(1)}$ and $a_{-1}^{(1)}$ were calculated
in~\cite{Rodrigo:2001jr} for the case where the observed photon is far 
from the collinear region. The extension of these results to the forward 
and backward regions is commented in Section~\ref{sec:untagged}.

Finally, the contribution from radiation of two real hard photons, 
$E_{\gamma}>w\sqrt{s}$, is calculated numerically using the 
helicity amplitude method~\cite{Jegerlehner:2000wu}.
The sum of the virtual plus soft corrections to one single photon 
events and the hard contribution of two photon emission gives the 
final result at NLO which is independent of the soft photon cutoff $w$.

\section{Monte Carlo simulation}

A Monte Carlo generator has been built which simulates the production 
of two charged pions together with one or two hard photons and includes 
virtual and soft photon corrections to the emission of one single real 
photon. It supersedes the previous versions of the 
EVA~\cite{Binner:1999bt,Rodrigo:2001jr} Monte Carlo.
Again the program exhibits a modular structure which preserves the 
factorization of the initial state QED corrections. The simulation of 
other exclusive hadronic channels can, therefore, be easily included with 
the simple replacement of the current(s) of the existing modes, and the 
corresponding multi-particle hadronic phase space. The simulation of 
the four pion channel~\cite{Czyz:2000wh} will be incorporated soon 
as well as other multi-hadron final states. Our results will be 
presented in~\cite{inpreparation}.

\section{Tagged or untagged photons}

\label{sec:untagged}
 
From the experimental point of view it would be much easier to 
perform the analysis with no lower photon angular boundary.
The cross section in the forward and backward regions grows very fast 
and therefore a small deviation in the determination of the photon 
angle could introduce a large error. It turns out that further 
advantages appear if only an upper cut on the photon angle is 
imposed. The cross section is thus larger.
For instance, the cross section for radiative events with 
$\theta_{\gamma} < 21^\circ$ and $55^\circ<\theta_{\pi} < 125^\circ$
is 4 times bigger than the corresponding 
cross section for $5^\circ<\theta_{\gamma} < 21^\circ$.
Furthermore, since FSR is isotropic with respect to 
the beam axis its contribution does not change much 
and, as a result, its relative importance with respect to ISR is 
much smaller (see lower line in figure~\ref{fig:eva}). All together may 
provide a better control of the systematics of the measurement 
and, therefore, a better determination of the hadronic cross section. 

It should be point out that in the forward and backward regions some 
corrections of order $m_e^2/s$, where $m_e$ is the electron mass, 
has to be taken into account even though $m_e^2/s$ is a small 
quantity. At LO, the full leptonic tensor is given by: 
\begin{align}
L_0^{\mu \nu} &= 
\frac{(4 \pi \alpha/s)^2}{q^4} \; \bigg[ \left( 
\frac{2 m^2 q^2(1-q^2)^2}{y_1^2 y_2^2}
- \frac{2 q^2+y_1^2+y_2^2}{y_1 y_2} \right) g^{\mu \nu} \non \\ & 
+ \left(\frac{8 m^2}{y_2^2} - \frac{4q^2}{y_1 y_2} \right) 
\frac{p_1^{\mu} p_1^{\nu}}{s} 
+ \left(\frac{8 m^2}{y_1^2} - \frac{4q^2}{y_1 y_2} \right) 
\frac{p_2^{\mu} p_2^{\nu}}{s} \non \\
& - \left( \frac{8 m^2}{y_1 y_2} \right) 
\frac{p_1^{\mu} p_2^{\nu} + p_1^{\nu} p_2^{\mu}}{s} \bigg]~, 
\label{Lmunu0}
\end{align}
with $y_i = 2 k_1 \cdot p_i/s$, $m^2=m_e^2/s$ and $q^2=Q^2/s$.
Expressing the bilinear products $y_i$ by the photon 
emission angle in the center of mass frame
\begin{equation*}
y_{1,2} = \frac{1-q^2}{2}(1 \mp \be \cos \theta)~, 
\qquad \be = \sqrt{1-4m^2}~.
\end{equation*}
The electron mass enters twofold. Through $y_i$, $i=1,2$,
but also quadratically in terms of the type $m^2/y_i^2$. In the former
case, the electron mass regulates the collinear divergences of the cross 
section. The second kind of contributions gives only finite corrections 
which, nevertheless, can be sizeable. In both cases the electron mass 
plays a relevant role only when the photon is emitted at very small angles.
Otherwise, the limit $m_e^2/s\rightarrow 0$ can be taken safely. 

\begin{figure}
\begin{center}
\epsfig{file=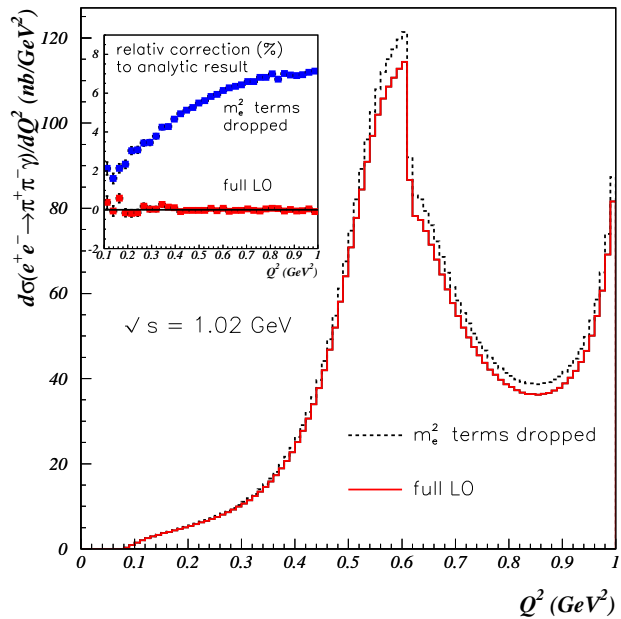,width=9cm}
\end{center}
\caption{Effect of the quadratic terms in the electron mass in the 
LO prediction for the $e^+ e^- \rightarrow \pi^+ \pi^- \gamma$ 
differential distribution. $E_{\gamma}>10$~MeV.}
\label{fig:less}
\end{figure}

\begin{figure}
\begin{center}
\epsfig{file=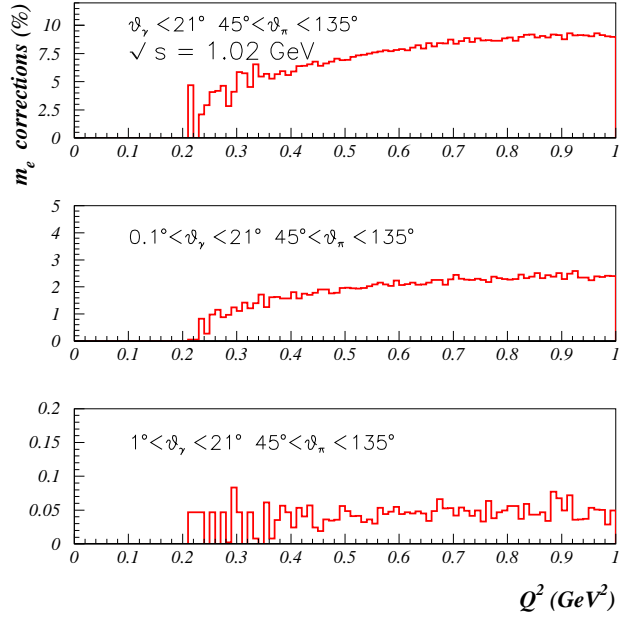,width=9cm}
\end{center}
\caption{Electron mass corrections for different photon angular cuts.}
\label{fig:smallangles}
\end{figure}

To illustrate this point the differential cross section for the process 
$e^+ e^- \rightarrow \pi^+ \pi^- \gamma$ at $\sqrt{s}=1.02$~GeV is 
presented in figure~\ref{fig:less}. The pion and photon angles are 
integrated without applying any angular cut. The electron mass is kept in the 
scalar bilinears $y_i$ but the quadratic terms are considered or not
to show their effect. Both results are compared in the small 
insert of this figure with the known corresponding analytical 
differential distribution 
\begin{align}
Q^2 \frac{d\sigma}{dQ^2} = \frac{4\alpha^3}{3 s} R(Q^2)
\bigg\{ \frac{s^2+Q^4}{s(s-Q^2)} 
\left( \log \frac{s}{m_e^2}-1 \right)
\bigg\}~,
\label{diff1}
\end{align}
with the substitution 
$R(Q^2) \rightarrow (1-4m_{\pi}^2/Q^2)^{3/2}$ $\mid~F_{2\pi}(Q^2)\mid^2/4$
for the two pion exclusive channel, being $F_{2\pi}$ the pion form factor. 
In (\ref{diff1}) the expansion in the electron mass has been performed after 
integration. 
Figure~\ref{fig:less} shows thus the good performance of the Monte Carlo 
integration also at very small photon angles. The analytical result agrees
with the Monte Carlo integration when the full electron mass dependence 
is considered in~\Eq{Lmunu0}. If the quadratic mass terms are not taken into 
account a deviation of up to $8\%$ is found. 
In contrast, at higher energies these corrections are quite suppressed, 
less than 1 per mil at $\sqrt{s}=10$~GeV.

Figure~\ref{fig:smallangles} shows the effect of the electron mass 
corrections for different angular cuts at $\sqrt{s}=1.02$~GeV. 
Up to $10\%$ deviation with respect to the full result is found which 
nevertheless decreases quite steeply if the collinear region is excluded:
to less than 5 per mil for tagged photon events at $1$ degree minimal angle. 

These electron mass finite corrections, which are relevant
only in the very forward and backward regions, below 1 degree at 
$\sqrt{s}=1.02$~GeV, have been included in the new version of 
the EVA Monte Carlo. At present, only at LO. At NLO, the corresponding 
electron mass corrections to the virtual and soft contributions in 
one photon events are included in the definition of the $y_i$'s
while the quadratic terms will be included soon.
For two photon events the full electron mass dependence 
is already considered.

\section{Conclusions}

The radiative return method is a competitive method to measure
the hadronic cross section at high luminosity $e^+ e^-$ colliders 
($\phi$ and $B$-factories) giving access to a wide range of energies, 
from threshold to the center of mass energy of the collider.
A Monte Carlo generator has been built which simulates the production 
of two charged pions together with one or two hard photons and includes 
virtual and soft photon corrections to the emission of one single real 
photon. Its modular structure is such that the simulation of other 
exclusive hadronic channels can be easily included. The description 
of events with hard photons radiated at very small angle
has been also investigated.


It is a pleasure to thank the organizers of this meeting for the 
stimulating atmosphere created during the school and the Katowice 
group for his kind hospitality. The author acknowledges J.~H.~K\"uhn
and H.~Czy\.z for a fruitful collaboration.
Work supported in part by BMBF under grant number 05HT9VKB0,
E.U. EURODAPHNE network TMR project FMRX-CT98-0169
and E.U. TMR grant HPMF-CT-2000-00989.



\begin{thebibliography}{99}

\bibitem{Brown:2001mg}
H.~N.~Brown {\it et al.}  [Muon g-2 Collaboration],
Phys.\ Rev.\ Lett.\  {\bf 86} (2001) 2227
[hep-ex/0102017].

\bibitem{Akhmetshin:1999uj}
R.~R.~Akhmetshin {\it et al.}  [CMD-2 Collaboration],
[hep-ex/9904027].

\bibitem{reviews}
J.~Prades,
[hep-ph/0108192].
K.~Melnikov,
[hep-ph/0105267].
W.~J.~Marciano and B.~L.~Roberts,
[hep-ph/0105056].

\bibitem{amuQED}
V.~W.~Hughes and T.~Kinoshita,
Rev.\ Mod.\ Phys.\  {\bf 71} (1999) S133.
A.~Czarnecki and W.~J.~Marciano,
Nucl.\ Phys.\ Proc.\ Suppl.\  {\bf 76} (1999) 245
[hep-ph/9810512].

\bibitem{amuweak}
T.~V.~Kukhto, E.~A.~Kuraev, Z.~K.~Silagadze and A.~Schiller,
Nucl.\ Phys.\ B {\bf 371} (1992) 567.
S.~Peris, M.~Perrottet and E.~de Rafael,
Phys.\ Lett.\ B {\bf 355} (1995) 523 [hep-ph/9505405].
A.~Czarnecki, B.~Krause and W.~J.~Marciano,
Phys.\ Rev.\ D {\bf 52} (1995) 2619 [hep-ph/9506256];
Phys.\ Rev.\ Lett.\  {\bf 76} (1996) 3267 [hep-ph/9512369].

\bibitem{Eidelman:1995ny}
S.~Eidelman and F.~Jegerlehner,
Z.\ Phys.\ C {\bf 67} (1995) 585
[hep-ph/9502298].

\bibitem{Brown:1996eq}
D.~H.~Brown and W.~A.~Worstell,
Phys.\ Rev.\ D {\bf 54} (1996) 3237
[hep-ph/9607319].

\bibitem{Davier:1998si}
M.~Davier and A.~H\"ocker,
Phys.\ Lett.\ B {\bf 435} (1998) 427
[hep-ph/9805470].

\bibitem{Narison:2001jt}
S.~Narison,
Phys.\ Lett.\ B {\bf 513} (2001) 53
[hep-ph/0103199].

\bibitem{Jegerlehner:2001wq}
F.~Jegerlehner,
[hep-ph/0104304].

\bibitem{DeTroconiz:2001wt}
J.~F.~De Troc\'oniz and F.~J.~Yndur\'ain,
[hep-ph/0106025].

\bibitem{amulight}
E.~de Rafael,
Phys.\ Lett.\ B {\bf 322} (1994) 239
[hep-ph/9311316].
J.~Bijnens, E.~Pallante and J.~Prades,
Nucl.\ Phys.\ B {\bf 474} (1996) 379
[hep-ph/9511388];
Phys.\ Rev.\ Lett.\  {\bf 75} (1995) 1447
[Erratum-ibid.\  {\bf 75} (1995) 3781]
[hep-ph/9505251].
M.~Hayakawa and T.~Kinoshita,
Phys.\ Rev.\ D {\bf 57} (1998) 465
[hep-ph/9708227].
M.~Hayakawa, T.~Kinoshita and A.~I.~Sanda,
Phys.\ Rev.\ D {\bf 54} (1996) 3137
[hep-ph/9601310];
Phys.\ Rev.\ Lett.\  {\bf 75} (1995) 790
[hep-ph/9503463].

\bibitem{Krause:1997rf}
B.~Krause,
Phys.\ Lett.\ B {\bf 390} (1997) 392
[hep-ph/9607259].

\bibitem{Steinhauser:1998rq}
M.~Steinhauser,
Phys.\ Lett.\ B {\bf 429} (1998) 158
[hep-ph/9803313].

\bibitem{top}
J.~H.~K\"uhn and M.~Steinhauser,
Phys.\ Lett.\ B {\bf 437} (1998) 425
[hep-ph/9802241].
K.~G.~Chetyrkin, J.~H.~K\"uhn and M.~Steinhauser,
Phys.\ Lett.\ B {\bf 371} (1996) 93
[hep-ph/9511430].

\bibitem{Burkhardt:2001xp}
H.~Burkhardt and B.~Pietrzyk,
LAPP-EXP-2001-03.

\bibitem{Binner:1999bt}
S.~Binner, J.~H.~K\"uhn and K.~Melnikov,
Phys.\ Lett.\  {\bf B459} (1999) 279
[hep-ph/9902399].
K.~Melnikov, F.~Nguyen, B.~Valeriani and G.~Venanzoni,
Phys.\ Lett.\  {\bf B477} (2000) 114 
[hep-ph/0001064].

\bibitem{Spagnolo:1999mt}
S.~Spagnolo,
Eur.\ Phys.\ J.\ C {\bf 6} (1999) 637.
V.~A.~Khoze, M.~I.~Konchatnij, N.~P.~Merenkov, G.~Pancheri, L.~Trentadue and O.~N.~Shekhovzova,
Eur.\ Phys.\ J.\ C {\bf 18} (2001) 481
[hep-ph/0003313].

\bibitem{Czyz:2000wh}
H.~Czy\.z and J.~H.~K\"uhn, 
Eur.\ Phys.\ J.\ C {\bf 18} (2001) 497
[hep-ph/0008262].

\bibitem{Hoefer:2001mx}
A.~H\"ofer, J.~Gluza and F.~Jegerlehner,
[hep-ph/0107154].

\bibitem{Adinolfi:2000fv}
A.~Aloisio {\it et al.}  [KLOE Collaboration],
[hep-ex/0107023].
A.~Denig {\it et al.}  [KLOE Collaboration],
eConf {\bf C010430} (2001) T07
[hep-ex/0106100].
M.~Adinolfi {\it et al.}  [KLOE Collaboration],
[hep-ex/0006036].

\bibitem{babar}
E.~P.~Solodov  [BABAR collaboration],
eConf {\bf C010430} (2001) T03
[hep-ex/0107027].

\bibitem{Passarino:1979jh}
G.~Passarino and M.~Veltman,
Nucl.\ Phys.\ B {\bf 160} (1979) 151.

\bibitem{Rodrigo:2001jr}
G.~Rodrigo, A.~Gehrmann-De Ridder, M.~Guilleaume and J.~H.~K\"uhn,
Eur.\ Phys.\ J.\ C {\bf DOI 10.1007/s100520100784}
[hep-ph/0106132].

\bibitem{inpreparation}
G.~Rodrigo, H.~Czy\.z, J.~H.~K\"uhn and M.~Szopa, in preparation.

\bibitem{Jegerlehner:2000wu}
F.~Jegerlehner and K.~Ko\l odziej,
Eur.\ Phys.\ J.\ C {\bf 12} (2000) 77
[hep-ph/9907229].
K.~Ko\l odziej and M.~Zra\l ek,
Phys.\ Rev.\ D {\bf 43} (1991) 3619.

\bibitem{newlight}
M.~Knecht, A.~Nyffeler, M.~Perrottet and E.~d.~Rafael,
[hep-ph/0111059].
M.~Knecht and A.~Nyffeler,
[hep-ph/0111058].

\end{thebibliography}
\end{document}